# A simulation-optimization framework for food supply chain network design to ensure food accessibility under uncertainty

**Running Title:** SCND to ensure food accessibility under uncertainty

Mengfei Chen[a], Mohamed Kharbeche[b], Mohamed Haouari[c], Weihong (Grace) Guo[a,*]

[a] *Department of Industrial and Systems Engineering, Rutgers University, 96 Frelinghuysen Rd, Piscataway, NJ 08854*

[b] *Qatar Transportation and Traffic Safety Studies Center, College of Engineering, Qatar University, P.O. Box 2713 Doha, State of Qatar*

[c] *Mechanical and Industrial Engineering Department, College of Engineering, Qatar University, P.O. Box 2713 Doha, State of Qatar*

## CRediT authorship contribution statement

**Mengfei Chen:** Conceptualization, Methodology, Formal analysis, Investigation, Visualization, Software, Data curation, Writing - Original draft. **Mohamed Kharbeche**: Project administration, Funding acquisition, Resources, Writing - review & editing. **Mohamed Haouari**: Resources, Writing - review & editing. **Weihong (Grace) Guo**: Conceptualization, Methodology, Formal analysis, Visualization, Supervision, Writing - review & editing.

## Declaration of competing interest

The authors declare that they have no known competing financial interests or personal relationships that could have appeared to influence the work reported in this paper.

## Data availability statement

Data used in this paper are available by contacting the corresponding author.

*Corresponding author:

*Email address:* wg152@soe.rutgers.edu (W. Guo)

# A simulation-optimization framework for food supply chain network design to ensure food accessibility under uncertainty

**Running Title:** SCND to ensure food accessibility under uncertainty

ABSTRACT

How to ensure accessibility to food and nutrition while food supply chains suffer from demand and supply uncertainties caused by disruptive forces such as the COVID-19 pandemic and natural disasters is an emerging and critical issue. Unstable access to food influences the level of nutrition that weakens the health and well-being of citizens. Therefore, a food accessibility evaluation index is proposed in this work to quantify how well nutrition needs are met. The proposed index is then embedded in a stochastic multi-objective mixed-integer optimization problem to determine the optimal supply chain design to maximize food accessibility and minimize cost. Considering uncertainty in demand and supply, the multi-objective problem is solved in a two-phase simulation-optimization framework in which Green Field Analysis is applied to determine the long-term, tactical decisions such as supply chain configuration, and then Monte Carlo simulation is performed iteratively to determine the short-term supply chain operations by solving a stochastic programming problem. A case study is conducted on the beef supply chain in Qatar. Pareto efficient solutions are validated in discrete event simulation to evaluate the performance of the designed supply chain in various realistic scenarios and provide recommendations for different decision-makers.

## 1. Introduction

Food supply chains play a critical role in providing basics and necessities to support human activities and behaviors. The food supply chain is an essential component in ensuring food security, as it encompasses the entire process of getting food from farms and manufacturers to the end consumers. A well-designed food supply chain can help ensure that food is available, accessible, and affordable to people. This is especially important in times of crisis, such as natural disasters or pandemics, where the supply chain can be disrupted and affect the availability of food. Over the years, many people all over the world have suffered from insecure access to food. Even in developed countries like the U.S., about 33.8 million people, or 10.2% of U.S. households, were food insecure at some time during 2021 (Coleman-Jensen et al., 2022). The COVID-19 outbreak in 2020 aggravated the problem even more. For example, according to Feeding America (2022), the number of children facing hunger in the US rose from around 10 million in 2019 to 12 million in 2020 due to the pandemic. Unstable access to food also leads to insufficient nutrition intake and thus weakens the physical health of people. It may further lead to societal insecurity and political instability. A recent example is the mass protests in Sri Lanka in July 2022 caused by the food crisis, which is a result of the geopolitical tension between Ukraine and Russia as 40% of Sri Lanka's food consumption is imported from Ukraine. Therefore, improving and securing access to food is an important and urgent task in both developing and developed countries. However, existing literature only provides political advice on the macro level. There is few specific, quantitative, or implementable method that complies with the policies. In addition, existing works focus on securing food resources (e.g., improving agricultural productivity), yet it is evitable that disasters and climate uncertainties can make food resources out of control. Therefore, it is vital to target food security by finding an optimal way to allocate the limited food resources. To fill the gap, this paper develops a supply chain network design approach to solve the problem. While food accessibility is affected by many factors, this paper focuses on the design of food supply chains considering how people reach food and how people afford food.

The quality of food products is known to decay rapidly during production, storage, and delivery. Thus, the unique design considerations in food supply chains include factors like food quality, safety, and freshness within a limited time, making the food supply chain more complex to manage and design. In this paper, a basic supply chain network is designed to utilize resources such as vehicles, warehouse facilities, transportation routes, and workers within the supply chain and to achieve several objectives to meet the basic supply chain requirements. Building upon that, we introduce an objective function on food accessibility, and therefore the supply chain network design (SCND) is modeled as a multi-objective optimization problem.

Consumers' increasing demand for healthy, nutritional food inspired us to stress the importance of food accessibility in SCND. This is echoed by the fact that many governments have posted guidelines or requirements on residents' nutrition levels from a regular diet, reiterating the importance of food accessibility. However, although most existing research in SCND has considered cost, environmental impact, and efficiency of the network, none of them elaborate on food accessibility. This research gap brings us to investigate SCND targeting food accessibility. Another gap in the literature is that there is no consensus on how to quantify food accessibility. Some studies measure food accessibility by resource availability, based on the ratio of food providers and population; some calculate the travel time from a residential area to the closest food provider; others consider nonspatial factors like race, income, educational attainment, and unemployment rate. Each of these quantitative measures is designed for its specific situation and thus cannot be generalized to food supply chains to guide design and operations. Moreover, the food nutrition levels are not considered in existing measures of food accessibility, motivating us to develop a novel food accessibility evaluation approach to be incorporated into the SCND. Meanwhile, the cost of supply chain operations also needs to be considered.

To fill in the gap between SCND and food accessibility, this paper aims to find the optimal design of food supply chains with two objectives, food accessibility and cost, respectively. Therefore, an optimization model with two objective functions is proposed. Since the optimal design involves both long-term, tactical decisions (e.g., locations) and short-term, operational decisions (e.g., inventory and replenishment), we propose two phases for the SCND. We adopt a single-channel supply pattern involving three echelons: warehouse/supplier, distribution center (DC), and customer.

In Phase I, we determine the supply chain configuration, including the locations of DCs, the linkages between warehouses and DCs, and the linkages between DCs and customers. Phase I is timely when an existing supply chain needs to be re-designed due to new highways or facilities, or when a new supply chain network is needed like when Qatar, a country heavily relies on global imported food, is now shifting to more domestic food production and so a new supply chain network needs to be designed and continuously optimized. The core of Phase I is a facility location problem solved as integer programs, in which we choose which facilities to open, which DCs to replenish from which warehouse, and which DCs to supply which customers, in order to satisfy some fixed demand at minimum cost.

The entire supply chain configuration from Phase I may be divided into several regions to account for different costs of living, income, and administrative policies in different demographic regions. In Phase II, we zoom in to each region to determine the supply chain operations, including the quantity of food distributed from each warehouse to each DC and from each DC to each customer during each time period, and the inventory level at the beginning of each time period, in order to maximize accessibility at minimum cost. These decision variables are continuous.

With the presence of both integer and continuous decision variables, the proposed two-phase framework is a stochastic multi-objective mixed-integer optimization problem with nonlinear constraints (terminology explained in Appendix A). Both demand and supply are modeled as stochastic variables. Uncertainty in the supply may be caused by factors such as disruptions and policy changes. Uncertainty in the demand may be caused by factors such as macroeconomic risks, competitive risks, and customers' changing preferences. The nonlinear constraints reflect the proposed food accessibility evaluation index.

The technical route of this paper is shown in Fig. 1. The objective is to find the optimal tactical (location and linkage) and operational decisions (inventory and distribution quantity) for a three-echelon supply chain network. The first echelon is a set of warehouses, representing factories or suppliers or ports/airports for imported food. The second echelon refers to a set of distribution centers (DCs), and the third echelon includes a set of customers or retailers. We will determine the locations of DCs, the linkages, and material flows between echelons. The objective functions include maximizing food accessibility and minimizing total cost. The total cost consists of three cost elements: transportation cost to account for the weighted shipping cost from DCs to customers, inventory cost that encompasses the expenses associated with stock levels in DCs, and unfulfilled demand cost since we want to satisfy customer demand as much as possible.

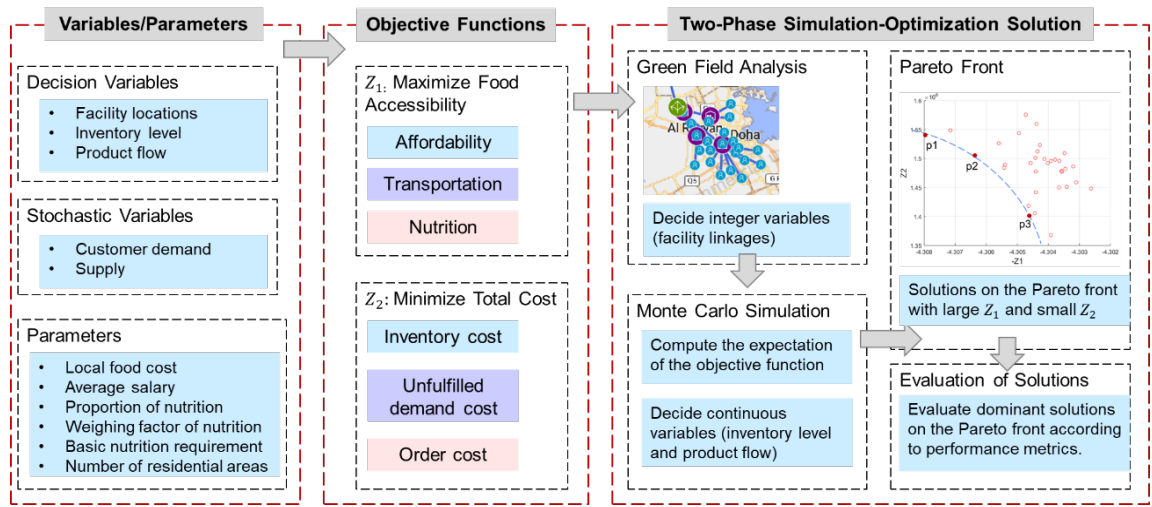

**Fig. 1.** Technical route of the proposed simulation-optimization framework for food supply chain network design to ensure food accessibility under uncertainty.

After formulating the stochastic multi-objective mixed-integer programming problem for SCND, we propose a simulation-optimization framework to solve it. A food accessibility evaluation index system is developed to measure local food accessibility. In Phase I of the proposed framework, the Green Field Analysis (GFA, explained in Appendix A) is applied to determine the optimal locations of DCs and the linkages between echelons. In Phase II, because of demand and supply uncertainty, a two-stage stochastic programming model is developed to maximize food accessibility at minimum total cost. Monte Carlo simulation (explained in Appendix A) is used to estimate the expected values of the objective functions. We iteratively perform a Monte Carlo simulation to obtain a set of feasible solutions. The Pareto efficient solutions (explained in Appendix A) on the Pareto front (explained in Appendix A) are then selected from all feasible solutions. Lastly, each selected Pareto solution is validated in discrete event simulation, demonstrating how the designed supply chain performs in a more realistic scenario and providing evidence for decision-making.

The remainder of this paper is divided into the following sections. Section 2 presents a review of the scientific studies that focused on the most recent efforts dealing with the SCND problem and existing food accessibility measurement criteria. Section 3 provides a detailed description of the SCND problem, including the evaluation index system proposed for food accessibility and the mathematical model formulated for the studied problem. Section 4 presents the proposed simulation-optimization solution approach to solve the studied problem. Section 5 demonstrates the proposed model and solution approach using a case study for the food supply chain in Qatar. Section 6 provides concluding remarks along with a set of future research extensions.

## 2. Literature review

In this section, we present a review of the related scientific studies with a focus on (1) ensuring food security, (2) food accessibility and evaluation criteria in SCND, and (3) solving stochastic SCND. Section 2.4 summarizes the literature review and highlights the contributions of this paper.

### 2.1. Ensuring food security

The majority of the literature addresses food security issues at the macro level by providing policy recommendations. Jaenicke and Virchow (2013) identify commonalities and parameters in developing successful nutrition-sensitive agriculture in different locations in order to improve food and nutrition security. Lawson-Lartego and Cohen (2020) come up with ten recommendations for African governments to ensure food and nutrition security for poor and vulnerable populations during diseases like COVID-19. The first thing they point out is to protect food supply chain and consider them essential services. Tanumihardjo et al. (2020) use the Sustainable Development Goals (SDGs) as a framework to illustrate the importance of maize agro-food systems for ensuring food and nutrition security. Popkova (2022) point out two potential solutions to food insecurity: reducing the demand for food and sustainable development of agriculture. Skripko and Bodrug (2022) stress the importance of the creation of innovative food to ensure food security. Howard et al. (2023) identify an opportunity for food security through supporting the expansion of inland fisheries by improving the existing preservation and storage techniques. Alam et al. (2024) explore the interaction of enablers and identify key factors that can mitigate the ripple effect in the global grain supply chain, but they have not formulated proper mitigation strategies. Most existing literature addressing food security stays at the policy level, without specifying exact actions. Therefore, it is necessary to develop more specific and implementable methods to comply with macro policies.

There are a few recent studies that focus on investigating models to ensure agricultural productivity. They aim at achieving food security by guaranteeing plenty of food recourses. Iskhakov et al. (2022) develop a regression model to predict agricultural productivity, but do not point out what further actions to be made to ensure food security. Talukdar et al. (2022) propose a method that ensures food security by efficient management of land and increasing food production. They design a framework that integrates optimization and damage estimation approach to determine the optimal allocation of sowing areas. However, agricultural productivity may not be as expected due to climate uncertainty and limited land resources. In addition, food security cannot be achieved without correctly and efficiently distributing food resources. This inspires us to take the food supply chain as an attempting way to solving the problem. Existing research on SCND mostly considers cost, environment, and efficiency of the network, but none of them stresses the importance of food security (Govindan et al., 2014; Mohebalizadehgashti et al., 2020; Gholami-Zanjani et al., 2021; Naeeni and Sabbaghi, 2022). However, catastrophes such as the COVID-19 pandemic and locust swarm outbreaks have posed a significant threat to global food supply chain systems, both local and international food supplies suffered from severe disruptions (Xu et al., 2021). This makes it challenging to achieve food security under limited food supplements, especially for resource-poor, food-importing countries such as Qatar (Ben Hassen et al., 2020a). Meanwhile, a recent survey indicates that COVID-19 has been making people shift toward healthier diets (Ben Hassen et al., 2020b), implying there is an increasing demand for nutrition level of food. Therefore, it is necessary to stress the importance of the accessibility of nutrition in the domestic food supply chain.

*2.2. SCND and food accessibility*

Several works have been reported on measuring food accessibility. The National Cancer Institute measures food accessibility from the resource availability perspective, defining food accessibility to be the ratio of the number of food providers and the population within the same geographic unit (Dai and Wang, 2011). This measure has been adopted by many studies (Bingham and Zhang, 1997; Helling and Sawicki, 2003; Austin et al., 2005; Moore et al., 2008; Raja et al., 2008), but they did not consider the spatial interaction between food providers and customers across residential boundaries. To address this issue, several works used the nearest neighbor method to compute the distance or travel time from a residential area to the closest food provider (Pearce et al., 2006; Smoyer-Tomic et al., 2006; Sharkey and Horel, 2008). In addition, some papers also consider nonspatial factors such as race, income, educational attainment, and unemployment rate (Donkin et al., 1999; Helling and Sawicki, 2003; Guy and David, 2004; Algert et al., 2006; Larsen and Gilliland, 2008; Raja et al., 2008; Grace et al., 2017). Although there is existing research on food accessibility, there is no consensus definition or measurement criteria.

Recently, Zhang and Mao (2019) evaluated food accessibility as the easiness for a population to reach supermarkets, grocery stores, or other sources of healthy and affordable food. Their measure of food accessibility mainly focuses on spatial/geographic access to healthy food. Huang and Tian (2019) measured food accessibility in rural China as the number of food stores selling fruits and vegetables in the village. Zhang and Mao (2019) measured food accessibility according to the spatial interactions among three local factors, including healthy food stores, population, and transportation networks. Hu et al. (2020) measured accessibility from customers' viewpoint in terms of perceived time and effort, which stresses the importance of time and transportation mode.

The COVID-19 pandemic has changed people's dietary habits for healthier food and thus people's demand for nutritional food increased worldwide (Ben Hassen et al., 2020b; Di Renzo et al., 2020; Bennett et al., 2021). In addition, governments around the world have shown concerns about nations' nutrition balance, including the World Food Program (WFP) (Ntambara and Chu, 2021), the UK government (Mehta, 2020), the Indian government (Amin et al., 2020) and so forth. Therefore, accessible nutrition levels must be considered and elaborated upon when evaluating food accessibility. In existing research, however, food accessibility mostly considers only distance and transportation, ignoring nutrition levels, and there is no consensus on how food accessibility should be quantified. These show limitations in existing research.

*2.3. SCND under uncertainty and simulation-based optimization*

Mixed-integer programming (MIP) and stochastic programming (SP) have shown effectiveness in solving supply chain-related problems under uncertainty. The discrete and non-convex properties of MIP, along with the need to capture uncertainty via SP, have raised theoretical and computational challenges. The combination of these two classes of models, called stochastic mixed-integer programming (SMIP), deserves further investigation.

A few works have analyzed SCND with strategic and tactical decisions simultaneously using two-stage stochastic models. MirHassani et al. (2000) proposed a two-stage model for multi-period capacity planning of SCND and used Bender's decomposition to solve the resulting stochastic integer program. Tsiakis et al. (2001) developed a two-stage stochastic programming model for SCND under demand uncertainty; their development includes a large-scale mixed-integer linear programming to determine the number, location, and capacity of warehouses and DCs, the transportation linkages, and the flows and production rates of materials. Alonso-Ayuso et al. (2003) presented a two-stage stochastic 0-1 modeling and a branch-and-fix heuristic algorithm for SCND under uncertainty, in order to determine the production topology, plant sizing, product selection, product allocation, and vendor selection. Santoso et al. (2005) integrated a sampling strategy with an accelerated Benders decomposition algorithm to efficiently compute high-quality solutions to large-scale stochastic supply chain design problems. Azaron et al. (2008) developed a multi-objective stochastic programming approach for supply chain design under uncertainty, in which demands, supplies, processing, transportation, shortage, and capacity expansion costs are all considered as uncertain parameters.

The abovementioned methods solve problems with linear objective functions and constraints. In many real-world applications, however, the objective function or constraints have a nonlinear relationship with decision variables in SCND problems. Existing works solve the problem with approximation or simplify the nonlinear relationship to linearity for ease of calculation. For example, You and Grossmann (2008) proposed a mixed-integer nonlinear programming (MINLP) model where the total replenishment cost (the objective function) is a square root function of the annual expected demand of the product in each DC (the decision variable) since the replenishment cost is calculated by the economic order quantity model (EOQ), and they reformulated the problem by relaxing integer variables to continuous variables. Kaya and Urek (2016) also proposed an MINLP model based on EOQ to determine the optimal locations of facilities, inventory levels, and product prices. To simplify the model, they approximate square root functions to piecewise linear functions. Although approximation methods reduce problem complexity, the solutions may suffer from accuracy loss. Owing to simulation's strength in mimicking reality, increasing attention has been devoted to solving complex optimization problems by simulation approaches.

Recently, simulation-based optimization has been an efficient and effective decision-making approach due to its ability to obtain high-fidelity optimal solutions that satisfy nonlinear objective functions and constraints. Yoo et al. (2010) applied a discrete-event simulation-based optimization approach for SCND. They combined global random search using nested partition and statistical selection using optimal computing budget allocation algorithm into a hybrid algorithm to solve the optimization problem. Salem and Haouari (2017) applied Monte Carlo simulation to solve a three-echelon SCND with random supply and random demand. Inspired by simulation's strength in mimicking complicated supply chain systems, we propose a hierarchical simulation-optimization framework, in which the integer variables for linkages between echelons are determined by GFA in Phase I, whereas continuous variables for inventory levels and material flows are solved with Monte Carlo simulation in Phase II.

*2.4. Literature summary and contributions*

A detailed review of the collected literature indicates that only a few studies have discussed the food accessibility concept in SCND without providing any supporting models for food accessibility evaluation methods that can be used by the relevant supply chain stakeholders. On the other hand, the literature dealing with stochastic mixed integer nonlinear programming problems is fairly broad. Several variants of solving SCND under uncertainty have been studied, including branch-and-fix heuristics, Benders decomposition algorithm, and simulation-based methods. Considering the existing trends in supply chains and potential benefits that can be offered by the food accessibility modeling to people's livelihood and nation's wellbeing, this study aims to provide the following contributions to the state-of-the-art:

- A specific and quantitative method to ensure food security by finding an optimal way to allocate the limited food resources.

- A novel mathematical formulation is proposed for food accessibility evaluation, aiming to quantify food accessibility from three dimensions (affordability, easiness of transportation, and quality).

- A multi-objective, mixed-integer stochastic programming model is formulated to find the optimal design of food supply chains to maximize food accessibility and minimize cost.

- The optimization problem aims to determine long-term, tactical decisions (e.g., locations) and short-term, operational decisions (e.g., inventory and replenishment), considering demand and supply uncertainty.

- Due to the problem's complexity, a two-phase hierarchical simulation-optimization framework consisting of GFA and Monte Carlo simulation is developed to solve the problem.

- A case study on the beef supply chain in Qatar is performed to assess the performance of the developed method in a real-world application, demonstrating the effectiveness and practical value of the proposed method.

- Optimal solutions are validated in discrete event simulation to evaluate how the designed supply chain performs in a more realistic scenario, providing decision support for stakeholders.

## 3. Problem description and modelling

*3.1. Problem description*

In this paper, we propose a simulation-optimization framework to solve an SCND problem under uncertainty. The purpose is to design a three-echelon supply chain network to achieve an optimal policy over $T$ consecutive periods. The three echelons include warehouses, DCs, and customers (retail stores). The designed supply chain configuration is divided into several regions because of different costs of living, income, and administrative policies in different demographic regions. We choose a single-channel supply pattern since it is one of the most typical and widely used supply patterns. Fig. 2 shows an example of a single-channel three-echelon SC. The material flow is from warehouses to DCs and then to customers. The single-channel supply pattern specifies that each node only receives from one upstream node.

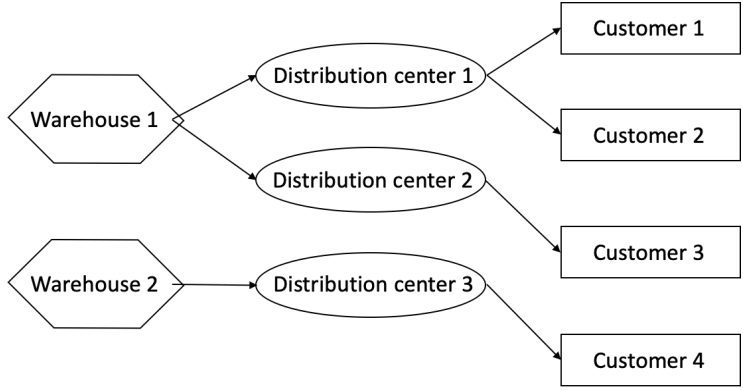

**Fig. 2.** An example of a single-channel three-echelon SC.

The notations for sets, parameters, and outcomes are presented in Table 1. We denote the set of regions as $I$, the set of warehouses as $W$, the set of DCs in region $i$ as $H_i$, and the set of customers in region $i$ as $L_i$. Considering the limited capacity of warehouses and DCs, we denote $Cap_w$ as the capacity of warehouse $w$, $Cap_h^{DC}$ as the capacity of the DC $h$ in region $i$. We also consider safety stock in DCs to prevent an out-of-stock situation against fluctuated supply and demand. We denote $v$ as a percentage parameter for the average local demand, where $0 \leq v \leq 1$, and $v \cdot S_h^{DC}$ ($h \in H_i$) as the safety stock, which reflects the minimum amount of a product stored in the DC. We denote $\tilde{\theta}_{wht}$ as the stochastic parameter from warehouse $w$ to DC $h$ at time $t$, which means if warehouse $w$ delivered $\sum_{w \in W} x_{wht}$ units, DC $h$ will receive $\sum_{w \in W} (1 - \tilde{\theta}_{wt}) \cdot x_{wht}$ units. We denote $\widetilde{D}_{lt}$ as the demand of customer $l$ at time $t$, which is a random variable.

**Table 1**

Notation for sets, parameters, and variables.

| Sets | | |
|---|---|---|
| | $I$ | Set of regions. |
| | $W$ | Set of warehouses. |
| | $H_i$ | Set of DCs in region $i$, $i \in I$. |
| | $L_i$ | Set of customers in region $i$, $i \in I$. |
| Parameters | | |
| $I_{it}^X \ (X = A, T, Q)$ | | Accessibility evaluation index for affordability, transportation, and nutrition level in time period $t$. |
| $c_{hlt}$ | | Quantity of food (kg) delivered from DC $h$ to costumer $l$ at time step $t$. |
| $l_i$ | | Local food cost in region $i, i \in I$. |
| $s_i$ | | Average income in region $i, i \in I$. |
| $\beta_j$ | | Amount of nutrition $j$ provided by the food item. |
| $q_j$ | | Weighing factor of nutrition $j$. |
| $r_j$ | | Minimum requirement of nutrition $j$ per person per day. |
| $p_i$ | | Number of residential areas $i, i \in I$. |
| $d_{hl}$ | | Distance between DC $h$ and customer $l$. |
| $f_{hl}$ | | Weighing factor of the path from DC $h$ to customer $l$. |
| $S_h^{DC}$ | | Average local demand in DC $h, h \in H_i$. |
| $v$ | | Percentage parameter in safety stock. |
| $\pi_h$ | | Inventory cost per unit in DC $h, h \in H_i$. |
| $u_i$ | | Unit cost of unfulfilled demand in region $i, i \in I$. |
| $o_{wh}$ | | Unit cost of orders from warehouse $w, w \in W$. |

**Stochastic parameters**

| | |
|---|---|
| $\widetilde{D}_{lt}$ | Demand from customer $l$ at time $t$. |
| $\tilde{\theta}_{wht}$ | Stochastic parameter from warehouse $w$ to DC $h$ at time $t$. |

**Phase I variables**

| | |
|---|---|
| $z_{wh}$ | Binary variable, $z_{wh} = 1$ if warehouse $w$ is assigned to DC $h$. It ensures the linkage between warehouses and DCs. |
| $y_{hl}$ | Binary variable, $y_{hl} = 1$ if DC $h$ is assigned to customer $l$. It ensures the linkage between DCs and customers. |

**Phase II variables**

| | |
|---|---|
| $\phi_1(z, y, \widetilde{D}, \tilde{\theta})$ | Functions of the decision variables $z_{wh}$, $y_{hl}$ and stochastic variables $\widetilde{D}_t$, $\tilde{\theta}_t$ in objective function $Z_1$. |
| $\phi_2(z, y, \widetilde{D}, \tilde{\theta})$ | Functions of the decision variables $z_{wh}$, $y_{hl}$ and stochastic variables $\widetilde{D}_t$, $\tilde{\theta}_t$ in objective function $Z_2$. |
| $x_{wht}$ | Quantity of food to be ordered by DC $h$ from warehouse $w$ in each time period $t$. |
| $c_{hlt}$ | Quantity of food to be delivered from DC $h$ to customer $l$ in each time period $t$. |
| $Inv_{h,t}^{DC}$ | Inventory level at the DC $h$ in time period $t$. |

Furthermore, we denote $r_j$ as the minimum requirement of nutrition $j$ per person per day, which is obtained from government guidelines, and $p_i$ as the number of residential areas in region $i$. Residential areas are communities where people live and gather. The number of residential areas is proportional to the local population. The weighting factor of nutrition $j$ ($q_j$) is introduced to reflect the relative importance of micronutrients. If there is no special consideration, we take $q_j = 1$. We denote $\pi_h$ as the inventory cost per unit of DC $h$, $u_i$ ($i \in I$) as the unit cost of unfulfilled demand in region $i$, and $o_{wh}$ as the unit cost of orders from the warehouse $w$ to DC $h$.

The objective functions are to maximize food accessibility and minimize the total cost. The cost includes the sum of inventory cost, unfulfilled demand cost, and order cost. In Phase I, the target is to determine tactical decisions for long-run periods, including:

1. The location and number of DCs in each region.
2. The linkage between warehouses and DCs.
3. The linkage between DCs and customers.

In Phase II, the target is to determine operational decisions, including:

1. The quantity to be ordered by each DC from each warehouse in each period.
2. The quantity to be delivered from each DC to each customer.
3. The inventory level for each DC in each period.

### 3.2. Food accessibility evaluation index system

Food accessibility will be measured using three dimensions. The three dimensions of the food accessibility evaluation index system are explained in Table 2.

**Table 2**

Three dimensions in the food accessibility evaluation index system.

| Notation | Dimension | Description |
|---|---|---|
| A | Affordability | How much to pay versus how much people earn. |
| T | Transportation | Easiness of transportation to food. |
| Q | Quality | Nutrition requirement versus available nutrition in food. |

First, the Affordability (A) dimension evaluates how much food the local people can purchase, which is measured by the local food price $l_i$ and average income $s_i$ in region $i$. Equation (1) shows the calculation of affordability, which is the ratio of local food cost to local income.

$$C_{it}^A = \frac{l_i}{s_i} \tag{1}$$

Easiness of transportation (T) evaluates the convenience and time cost to reach food. If it takes too much effort, people may be less willing to purchase food. The transportation effort is related to the distance ($d_{hl}$) between DC $h$ and customer $l$, as well as the quantity of food purchased ($c_{hlt}$) from DC $h$ by customer $l$ during time $t$. Equation (2) calculates the transportation convenience.

$$C_{it}^T = \sum_{h \in H_i} \sum_{l \in L_i} y_{hl} f_{hl} d_{hl} c_{hlt} \tag{2}$$

Quality (Q) refers to the nutrition level that people can obtain from the purchased food. This dimension helps to measure to what degree people can meet the nutrition requirement from the government. It is related to accessible nutrition $n_{ijt}$, the minimum requirement of nutrition $r_j$, and the population (measured by the number of residential areas $r_j$). The accessibility to nutrition should meet the minimum nutrition requirements as much as possible. The index of food quality and nutrition is measured by the deficiency of nutrition (the difference between the accessible nutrition and the required nutrition level), which is represented by Equation (3).

$$C_{it}^Q = \sum_{j \in J} q_j \cdot (n_{ijt} - r_j p_i)^+ \tag{3}$$

We denote $(n_{ijt} - r_j p_i)^+$ as the degree that the accessible nutrition can meet the requirement. The minimum requirement $r_j p_i$ considers the nutrition requirement for each resident and the number of residents. If accessible nutrition $n_{ijt}$ is less than the requirement $r_j p_i$, $(n_{ijt} - r_j p_i)^+ = 0$, otherwise $(n_{ijt} - r_j p_i)^+ = n_{ijt} - r_j p_i$. The calculation of $n_{ijt}$ is represented by Equation (4). Accessible nutrition is related to the total local inventory of related food and the amount of nutrition contained in food.

$$n_{ijt} = \sum_{h \in H_i} Inv_{h,t}^{DC} \cdot \beta_j \tag{4}$$

Since parameters in three indices have different units, we use Equation (5) to do normalization, so that the three indices have the same scale.

$$I_{it}^X = \frac{C_{it}^X}{C_{it}^X - C_1^X} \quad (X = A, T, Q) \tag{5}$$

where $C_{it}^X$ is the calculation for each evaluation index. $C_1^X$ is the value used to scale $I_i^X$ to be within $[0,1]$. Residents' affordability of food from an economic aspect can be measured by how much to pay versus how much they can earn.

## 3.3. Mathematical modeling

In this section, we describe the proposed mathematical formulation. In Phase I, we determine the long-run decision variables, including the binary variables $z_{wh}$ and $y_{hl}$ which indicate the linkages between warehouses, DCs, and customers. In Phase II, we determine the operational decision variables, including the distribution quantities in each time period $x_{wht}$ and $c_{hlt}$, and the inventory level $Inv_{h,t}^{DC}$. Since customer demand and supply from the warehouse are uncertain, we formulate the problem as a two-stage stochastic model. Decision variables are explained in detail in Table 1.

### 3.3.1. Formulation of Phase I

In Phase I, we make long-term decisions using the Green Field Analysis (GFA) approach. The objectives may change across time, and thus for the long-term decision, we only consider the distance. The objective of the GFA is to determine the best location for DCs. The purpose is to find the location that allows us to fulfill our customer demands at the lowest total transportation cost based on the Gravity model (Ivanov et al., 2019).

We denote $(a_l, b_l), l \in L_i, i \in I$ as the coordinate of a customer. In addition, we denote $(p_h, q_h), h \in H_i, i \in I$ as the coordinate of the DC. $\mu_l$ is denoted as the average demand of customer $l$. We also assume the transportation cost is linearly proportional to the distance and the transportation volume (the demand). We have the objective function represented by Equation (6). The model is called the center-of-gravity model of location analysis. $d\big((p_h, q_h); (a_l, b_l)\big)$ is the distance between $(a_l, b_l)$ and $(p_h, q_h)$.

$$W = \text{Minimize} \sum_{l \in L_i} d\big((p_h, q_h); (a_l, b_l)\big) \cdot \mu_l \qquad (6)$$

### 3.3.2. Formulation of Phase II

In Phase II, continuous variables $x_{wht}$, $c_{hlt}$ and $Inv_{h,t}^{DC}$ are determined. We formulate the problem as a stochastic programming model where the first-stage model calculates the objective function based on the expectation of terms $\phi_1(z, y, \widetilde{D}, \widetilde{\theta})$ and $\phi_2(z, y, \widetilde{D}, \widetilde{\theta})$ relevant to integer decision variables and stochastic variables, and the second-stage model solves for $\phi_1(z, y, \widetilde{D}, \widetilde{\theta})$ and $\phi_2(z, y, \widetilde{D}, \widetilde{\theta})$.

1) Formulation of the first-stage model:

$Z_1$ is the objective function maximizing the expectation of food accessibility, which is measured by the evaluation index system proposed in Section 3.2. $w_i^A I_{it}^A$ has no relation with stochastic variables, and so it can be taken outside the expectation. Equation (7) indicates the formulation of $Z_1$. $Z_2$ is the objective function minimizing the expectation of total cost, which is shown in Equation (8). $\phi_1(z,y,\widetilde{D},\widetilde{\theta})$ and $\phi_2(z,y,\widetilde{D},\widetilde{\theta})$ describes functions of the decision variables $z_{wh}$, $y_{hl}$ and stochastic variables $\widetilde{D}_t$, $\widetilde{\theta}_t$, which will be discussed in Phase II.

$$Z_1 = \text{Maximize} \sum_i \sum_{t=1}^T \{w_i^A I_{it}^A + E_{\widetilde{D},\widetilde{\theta}}[\phi_1(z,y,\widetilde{D},\widetilde{\theta})]\} \qquad (7)$$

$$Z_2 = \text{Minimize} \sum_i \sum_{t=1}^T E_{\widetilde{D},\widetilde{\theta}}[\phi_2(z,y,\widetilde{D},\widetilde{\theta})] \qquad (8)$$

2) Formulation of the second-stage model:

We denote $\phi_1(z,y,\widetilde{D},\widetilde{\theta})$ and $\phi_2(z,y,\widetilde{D},\widetilde{\theta})$ as functions related to the decision variables $z_{wh}$, $y_{hl}$, and stochastic variables $\widetilde{D}_t$, $\widetilde{\theta}_t$. The expression for $\phi_1(z,y,\widetilde{D},\widetilde{\theta})$ and $\phi_2(z,y,\widetilde{D},\widetilde{\theta})$ are given by Equation (9) and Equation (10), respectively. We see from (10) that this cost includes the inventory cost, the transportation cost, and the unfulfilled demand cost. $I_{it}^T$ and $I_{it}^Q$ are calculated by Equations (1)-(5).

$$\phi_1(z,y,\widetilde{D},\widetilde{\theta}) = \text{Maximize} \sum_i (-w_i^T I_{it}^T + w_i^Q I_{it}^Q) \qquad (9)$$

$$\phi_2(z,y,\widetilde{D},\widetilde{\theta}) = \text{Minimize} \sum_i \sum_{t=1}^T \sum_{h \in H_i} \pi_h \cdot Inv_{h,t}^{DC} + \sum_i \sum_{t=1}^T \sum_{l \in L_i} \sum_{h \in H_i} y_{hl} \cdot \rho_i \cdot g_{hlt} + \sum_{t=1}^T \sum_{h \in H_i} \sum_{w \in W} o_{wh} \cdot x_{wht} \qquad (10)$$

Subject to:

$$Inv_{h,t}^{DC} = Inv_{h,t-1}^{DC} - \sum_{l \in L_i} y_{hl} \cdot c_{hlt} + \sum_{w \in W} z_{wh} \cdot (1 - \widetilde{\theta}_{wt}) \cdot x_{wht} \qquad (11)$$

$$v \cdot S_h^{DC} \leq Inv_{h,t}^{DC} \leq Cap_h^{DC}, h \in H_i, l \in L_i \qquad (12)$$

$$\sum_{h \in H_i} z_{wh} \cdot x_{wht} \leq Cap_w, \ w \in W \qquad (13)$$

$$\sum_{h \in H_i} (g_{hlt} + c_{hlt}) = \widetilde{D}_{lt}, t = 0,1,2 \dots, T \qquad (14)$$

$$x_{wht} \geq 0, w \in W, h \in H_i, l \in L_i \qquad (15)$$

$$y_{hl} \in \{0,1\}, h \in H_i, l \in L_i \qquad (16)$$

$$z_{wh} \in \{0,1\}, w \in W, h \in H_i \qquad (17)$$

Constraint (11) represents the inventory balance between consecutive time periods. For each time period, the current inventory level equals the inventory level in the previous time period minus the output flow (total amount of food distributed to customers) and then plus the input flow (amount of food delivered from the warehouse).

To cope with demand and supply uncertainty, we place safety stock for each DC. The safety stock is designed to be proportional to the total local demand covered by each DC. We denote $v \in [0,1]$ as the proportion of local demand that will be used to decide the safety stock. $v$ is a preliminary decided value. In addition, the inventory level should not exceed the capacity, and thus the constraint for each DC can be represented by Constraint (12). Similarly, the total amount of food delivered from a warehouse during a time period should not exceed its capacity, as is shown by Constraint (13). Constraint (14) means in each time period, the customer demand equals the unfulfilled demand plus the distribution amount to the customer. Constraints (15)-(17) specify the variable types.

## 4. Simulation-optimization solution approach

### 4.1. Phase I solution

The objective function in Equation (7) is continuous and differentiable and the decision variables $p_h, q_h$ are unrestricted. Therefore, we can determine the optimal solution by differential calculus. We have:

$$\frac{dW}{dp_h} = \frac{|L_i| \cdot p_h}{\sqrt{(a_l - p_h)^2 + (b_l - q_h)^2}} - \sum_{l \in L_i} \frac{x_l}{\sqrt{(a_l - p_h)^2 + (b_l - q_h)^2}} \tag{18}$$

$$\frac{dW}{dq_h} = \frac{|L_i| \cdot q_h}{\sqrt{(a_l - q_h)^2 + (b_l - q_h)^2}} - \sum_{l \in L_i} \frac{y_l}{\sqrt{(a_l - p_h)^2 + (b_l - q_h)^2}} \tag{19}$$

Using demand data, Equations (18) and (19) are used to calculate the optimal coordinates of the DC. Setting the partial derivative formulas equal to 0, we obtain the solution in Equations (20) and (21).

$$p_h = \frac{\sum_{l \in L_i} \frac{\mu_l \cdot a_l}{\sqrt{(a_l - p_h)^2 + (b_l - q_h)^2}}}{\sum_{l \in L_i} \frac{\mu_l}{\sqrt{(a_l - p_h)^2 + (b_l - q_h)^2}}} \tag{20}$$

$$q_h = \frac{\sum_{l \in L_i} \frac{\mu_l \cdot b_l}{\sqrt{(a_l - p_h)^2 + (b_l - q_h)^2}}}{\sum_{l \in L_i} \frac{\mu_l}{\sqrt{(a_l - p_h)^2 + (b_l - q_h)^2}}} \tag{21}$$

The coordinates of the DCs have been decided. Based on this, the distance between DCs and customers can be calculated by Equation (22) accordingly.

$$d_{hl} = d\big((p_h, q_h); (a_l, b_l)\big) = \sqrt{(p_h - a_l)^2 + (q_h - b_l)^2}, h \in H_i, l \in L_i, i \in I \tag{22}$$

The linkages between customers and DCs are decided according to the minimum distance, and thus we obtained $z_{wh}$ and $y_{hl}$, $h \in H_i, l \in L_i, i \in I$ in the model.

*4.2. Phase II solution: Pareto method*

Since we have a multi-objective optimization (MOO) problem with unknown weights for each objective function, we use the Pareto method to obtain several optimal solutions. This method is used if the desired solutions produce a compromise solution (trade-off) and can be displayed in the form of Pareto optimal front (POF). To obtain the optimal front, we consider $Z$ as the linear combination of $Z_1$ and $Z_2$ with a linear factor $\varepsilon$. Let $Z = Z_1 - \varepsilon \cdot Z_2$, we solve by maximizing $Z$ by changing the value of $\varepsilon$ to obtain several feasible solutions. Fig. 3 shows an example of the solutions to a bi-objective problem. The dominated solutions are shown as the black points. Non-dominated solutions on the POF are shown as red stars. The Pareto optimal solutions and optimal value in MOO are usually achieved when one objective function cannot increase without reducing the other objective function. This condition is called Pareto optimality. The set of optimal solutions in MOO is called Pareto optimal solution.

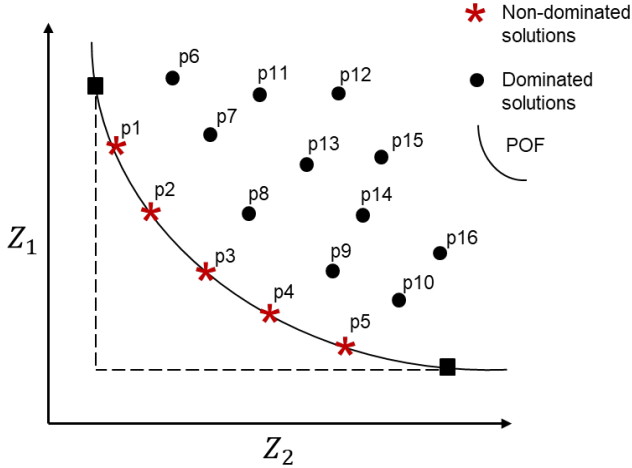

**Fig. 3.** An example of the Pareto optimal front.

For each value of $\varepsilon$, we solve for the model and obtain the corresponding optimal solution $(Z_1^{(r)}, Z_2^{(r)})$, $r = 1, 2 \ldots, R$, $R$ is the total number of solutions. The Pareto optimal front in our model is the set of optimal solutions that offer the highest accessibility for a defined level of the total cost. Define the set $U_z \coloneqq \{g: Z_2^{(g)} = z\}$ as the set of indices for a given level of $Z_2$, where $z \in \left[ \min_{r=1,2\ldots,R} Z_2^{(r)}, \max_{r=1,2\ldots,R} Z_2^{(r)} \right]$. For each given value of $z$, the index of the highest accessibility can be represented as Equation (23).

$$r_s^* = \underset{g \in U_Z}{\mathrm{argmax}}\, Z_1^{(g)}, s = 1,2,3 \ldots \tag{23}$$

Denote $S$ as the Pareto optimal front set, which can be represented by Equation (24).

$$S := \left\{ \left(Z_1^{(r_1^*)}, Z_2^{(r_1^*)}\right), \left(Z_1^{(r_2^*)}, Z_2^{(r_2^*)}\right), \ldots, \left(Z_1^{(r_s^*)}, Z_2^{(r_s^*)}\right), \ldots \right\} \tag{24}$$

After obtaining Pareto solutions and POF, we are going to put these solutions into a simulation model, because simulation can mimic the randomness of the supply chain system and the result is closer to reality. Through simulation, we can take a look at some index of interest (e.g., the service level, the unfulfillment demand, the order cost.) and discuss with stakeholders or ask for expertise. They may select a solution according to the aspect that they care about most.

*4.3. Phase II solution: Monte Carlo procedure*

Since there are stochastic variables representing the uncertainty of demand and supply, we use Monte Carlo simulation to solve the stochastic programming problem. In each replication, we randomly generate a sample of demand $\widetilde{D}^s$ and supply parameter $(1 - \widetilde{\theta}^s)$, and we then solve for the model in each time period. Finally, we compute the expectation of the objective function.

- Step 0: Set $s = 1$. Repeat Steps 1 and 2 $N$ times ($N$ is the number of replicates used to get an estimation of the accessibility).

- Step 1: Randomly generate a realization of uncertain variables $\widetilde{D}^s$ and $(1 - \widetilde{\theta}^s)$ with distribution $N(\mu, \sigma^2)$ and $U(a,b)$ respectively.

- Step 2: For each period $t$ ($t = 1, \ldots, T$) solve the following mathematical model:

$$\phi_t^s = \mathrm{Maximum} \sum_i \left(-w_i^T I_{it}^T + w_i^Q I_{it}^Q\right) - \\ \varepsilon \cdot \left( \sum_i \sum_{h \in H_i} \pi_h \cdot Inv_{h,t}^{DC} + \sum_i \sum_{l \in L_i} \sum_{l \in L_i} y_{hl} \cdot \rho_i \cdot g_{hlt} + \sum_{h \in H_i} \sum_{w \in W} o_{wh} \cdot x_{wht} \right) \tag{25}$$

Subject to:

$$Inv_{h,t}^{DC} = Inv_{h,t-1}^{DC} - \sum_{l \in L_i} y_{lh} \cdot \widetilde{D}_{lt} + \sum_{w \in W} z_{wh} \cdot (1 - \widetilde{\theta}_{wt}) \cdot x_{wht} \tag{26}$$

$$v \cdot S_h^{DC} \leq Inv_{h,t}^{DC} \leq Cap_h^{DC}, h \in H_i, l \in L_i \tag{27}$$

$$\sum_{h \in H_i} z_{wh} \cdot x_{wht} \leq Cap_w, \ w \in W \tag{28}$$

$$\sum_{h \in H_i} (g_{hlt} + c_{hlt}) = \widetilde{D}_{lt}, t = 0,1,2 \ldots, T \tag{29}$$

$$x_{wht} \geq 0, w \in W, h \in H_i, l \in L_i \tag{30}$$

$$y_{hl} \in \{0,1\}, h \in H_i, l \in L_i \tag{31}$$

$$z_{wh} \in \{0,1\}, w \in W, h \in H_i \tag{32}$$

Constraints (26)-(29) have the same interpretation as constraints (11)-(14) but are specifically formulated for each time period.

- Step 3: Compute the second-stage function $\varphi^s = \sum_{t=1}^{T} \varphi_t^s$.
- Step 4: Compute the estimate of the objective function $w_i^A I_i^A + \frac{1}{N}\sum_{s=1}^{N} \varphi^s$.

For each value of $\varepsilon$, we use the Monte Carlo simulation method to obtain the corresponding solutions. The output is a pair of optimal solutions $\left(Z_1^{(r)}, Z_2^{(r)}\right)$ at current $\varepsilon$ ( $r = 1, 2 \ldots, R$, $R$ is the total number of solutions). The solution provides the distribution amount from DC $h$ to customer $l$ ($c_{hlt}$) for each period, as well as the distribution amount from warehouse $w$ to DC $h$ ($x_{wht}$) for each period. In Section 5, we will demonstrate the solution procedure in a case study.

## 5. Case study

Based on the proposed framework, a numerical experiment is conducted for the State of Qatar's beef supply chain. Qatar used to be highly reliant on imported food and is now planning to stress more importance of self-production. The government constructs new facilities to support the development of the domestic supply chain. Because of the disruption caused by COVID-19 pandemic, supply and demand are faced with fluctuation and uncertainty. In the case study, we take the beef supply chain in Qatar as an example, because beef is one of the domestic products in Qatar. In this section, we first describe the parameters used in the case study, and then we discuss the results.

*5.1. Parameter setup*

This subsection describes how parameters in simulation and optimization models are identified from historical data. Local food costs and local income are obtained from public sources (Average Salary Survey, 2022; Index Mundi, 2022). We assume that 50,000 people form a residential area so that the number of residential areas in each region can be calculated accordingly. Table 3 summarizes local food costs, average local income, and residential areas in the four main regions in Qatar. The central region (Region 1) is the most populated, with a higher average income, while remote regions (Regions 2, 3, 4) are less populated, with a lower average income.

This disparity between different regions of the country also explains why the proposed method analyzes the supply chain for each region.

**Table 3**

Local food cost, local income, and residential areas (Average Salary Survey, 2022; Index Mundi, 2022).

| Region | Local food cost/kg ($l_i$) | Local income/year ($s_i$) | Population | Number of residential areas ($p_i$) |
|---|---|---|---|---|
| 1 | 21.77 QAR | 275,626 QAR | 1,114,595 | 22 |
| 2 | 21.77 QAR | 241,036 QAR | 235,434 | 5 |
| 3 | 21.77 QAR | 241,036 QAR | 105,853 | 2 |
| 4 | 21.77 QAR | 241,036 QAR | 218,583 | 4 |

Each vitamin and mineral have a specific role in the body. They are vital for immune function, brain development, and guaranteeing power needs, and thus an adequate intake of all micronutrients is necessary for optimal health. Table 4 summarizes the minimum daily micronutrient consumption requirement, as well as the constitution of the most critical micronutrients per 100g of beef (Hor, 2021). For example, the minimum daily Zn requirement is 2mg per person; 100g of beef contains 2.2mg Zn on average.

**Table 4**

Micronutrients requirement per day per person and the amount provided in 100g of beef (Hor 2021).

| Micronutrient | Minimum requirement | Amount provided in 100g of beef |
|---|---|---|
| Zn (mg) | 2 | 2.2 |
| Ca (mg) | 200 | 24 |
| Fe (mg) | 0.73 | 2.5 |
| A (μg) | 400 | 66 |
| B1 (μg) | 0.2 | 0.1 |
| B2 (mg) | 0.3 | 0.2 |
| B3 (mg) | 4 | 5.6 |
| B6 (mg) | 0.1 | 0.5 |
| B9 (μg) | 65 | 8 |
| B12 (μg) | 0.4 | 1.3 |
| C (mg) | 40 | 0.5 |
| D (μg) | 10 | 3 |

Table 5 shows the proportion of meat consumption in Qatar, based on a survey paper by Al-Thani et al. (2017). The percentage of beef (5%) and beef meat products (7%) totals 12% of all animal meat products. The daily beef consumption by a household is 33.6g, including 13.6g of beef and 20g of beef products (shown in bold). With the assumption of three people forming a household, the daily beef consumption per person is 33.6/3=11.2 (g/person).

**Table 5**

Animal meat products purchased in the State of Qatar according to nationality (2012-2013) (Al-Thani et al., 2017).

|  | Qatari | | Non-Qatari | | Total population (Households) | |
| --- | --- | --- | --- | --- | --- | --- |
|  | Amount*(g) | (%) | Amount*(g) | (%) | Amount*(g) | (%) |
| **Total animal meat group** | 343 | 100 | 182 | 100 | 287 | 100 |
| **Lamb** | 109.8 | 32 | 48.5 | 27 | 87.5 | 30 |
| **Beef** | 15.1 | 4 | 10.9 | 6 | **13.6** | **5** |
| **Beef meat products** | 23.1 | 7 | 13.9 | 8 | **20** | **7** |
| **Liver** | 3.2 | 1 | 2.7 | 1 | 2.9 | 1 |
| **Poultry** | 128.7 | 38 | 65.6 | 36 | 107.3 | 37 |
| **Fish** | 52.5 | 15 | 37.7 | 21 | 48 | 17 |
| **Other seafood** | 10.6 | 3 | 2.7 | 1 | 7.7 | 3 |

*(g/capita/day)

**Table 6**

Weighing factor $q_j$, minimum requirement $r_j$, and nutrition provided in 1kg of beef $\beta_j$.

| Micronutrient ($j$) | | Weighing factor ($q_j$) | Minimum requirement per day per person | Amount provided by beef to meet minimum requirement ($r_j$) | Amount provided in 1kg of beef ($\beta_j$) |
| --- | --- | --- | --- | --- | --- |
| 1 | Zn (mg) | 1 | 2 | 0.24 | 22 |
| 2 | Ca (mg) | 1 | 200 | 24 | 240 |
| 3 | Fe (mg) | 1 | 0.73 | 0.0876 | 25 |
| 4 | A (μg) | 1 | 400 | 48 | 660 |
| 5 | B1 (μg) | 1 | 0.2 | 0.024 | 1 |
| 6 | B2 (mg) | 1 | 0.3 | 0.036 | 2 |
| 7 | B3 (mg) | 1 | 4 | 0.48 | 56 |
| 8 | B6 (mg) | 1 | 0.1 | 0.012 | 5 |
| 9 | B9 (ug) | 1 | 65 | 7.8 | 80 |
| 10 | B12 (mg) | 1 | 0.4 | 0.048 | 13 |
| 11 | C (mg) | 1 | 40 | 4.8 | 5 |
| 12 | D (μg) | 1 | 10 | 1.2 | 30 |

Table 6 lists the parameters used in our model. They are obtained or calculated as follows. A total of 12 micronutrients are considered, same as those listed in Table 4. Since the government does not stress any specific nutrient, we assume they are of equal weights, $q_j = 1$ for all $j$'s. We assume that most of the required nutrition is provided by meat. Since beef consumption takes up 12% of the total meat consumption, we assert that beef should provide 12% of the required micronutrients. The amount of micronutrient $j$ must be provided by beef ($r_j$) in order to meet the minimum requirement can be calculated by multiplying the minimum requirement of micronutrition $j$ per kg (from Table 4) with 12%. The amount of micronutrient $j$ contained in 1 kg of beef ($\beta_j$) is ten times the amount per 100g from Table 4. Take Zn for example. The minimum requirement of daily Zn consumption is 2mg per person, and so the amount of Zn must be provided by beef to meet this requirement is $r_1$ = 2mg × 12% = 0.24mg; $\beta_1$ = 22mg since 100g of beef contains 2.2mg Zn on average.

## 5.2. Experimental results

### 5.2.1. Green Field Analysis (GFA)

We refer to the principal dairy farms in Qatar to obtain the locations of suppliers/warehouses, and thus their locations are fixed. Customers' locations are randomly generated within each region, and the number of customers in each region is proportional to the population listed in Table 3. DCs' locations and linkages are optimized in GFA from random initial locations. The initial setup is shown in **Fig. 4**(a), and the optimized configuration is shown in **Fig. 4**(b). Three warehouses with fixed locations provide beef for the four regions. The optimized configuration consists of four DCs in Region 1 (Doha), one DC in Region 2 (Al Khor), one DC in Region 3 (Al Daayen), and two DCs in Region 4 (Al Rayyan). The optimal linkages are shown in blue lines.

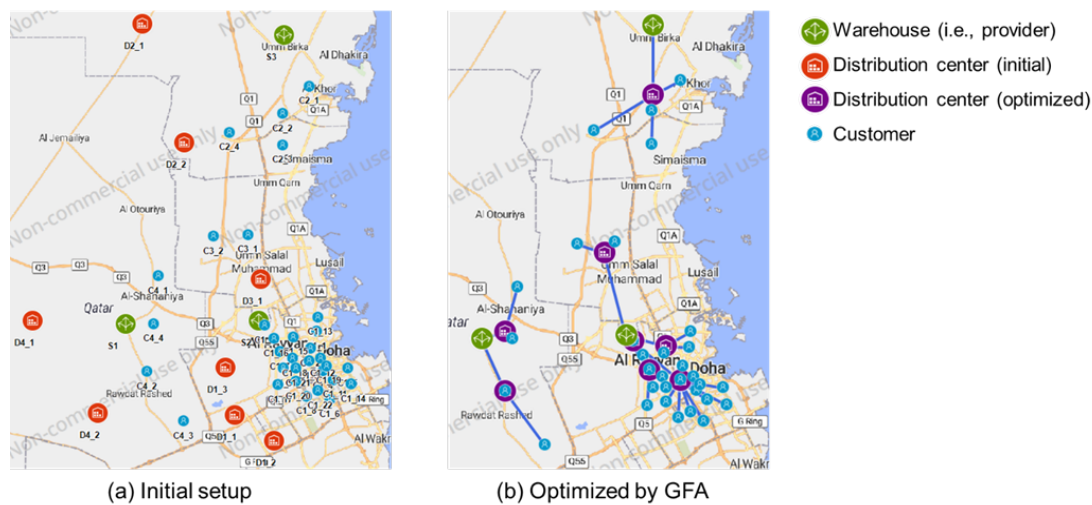

**Fig. 4.** Experiment setup for warehouse, DC, and customer: (a) Initial setup; (b) Optimized by GFA.

*5.2.2. Monte Carlo simulation*

In our Monte-Carlo simulation, we assume that customer demand has a normal distribution $N(\mu, \sigma^2)$ with $\mu = 560$ and $\sigma^2 = 50$. We assume the supply parameter $(1 - \tilde{\theta}_{wt})$ has uniform distribution $U(a, b)$ with $a = 0.8$ and $b = 0.9$. The unit cost of unfulfilled demand per day is $\rho_i = 5, i = 1,2,3,4$. The inventory cost per unit per day of DC $h$ is $\pi_h = 10, h \in H_i$. The unit cost of orders from warehouse $w$ per day is $o_{wh} = 3, w \in W, h \in H_i$.

Safety stock is considered to prevent an out-of-stock situation against fluctuated supply and demand, but it will also take up space and add to inventory costs. To discuss the trade-off of setting safety stock, we conduct two experiments to compare the results with safety stock and without safety stock. in the first experiment, we do not consider safety stock, so $v = 0$.

Solutions $\left(Z_1^{(r)}, Z_2^{(r)}\right)$ are shown as the circles in Fig. 5. Solutions at the lower left edge make up POF and are marked with red solid circles. The dashed line indicates the POF. Fig. 5(a) illustrates food accessibility versus total cost, and Fig. 5(b)-(d) indicates food accessibility versus inventory cost, unfulfilled demand cost, and order cost, respectively. Three Pareto solutions are found. The first solution (p1) leads to the highest food accessibility but has the highest inventory cost, unfulfilled demand cost, and order cost compared with the other two solutions. The second solution (p2) has moderate food accessibility and total cost, and it leads to the least unfulfilled demand cost. The last solution (p3) has the least cost but cannot achieve very high food accessibility compared with the other two solutions. The optimal solution can be determined according to stakeholders' preferences. If food accessibility is the primary thing to consider, we can choose the first solution. If the fulfillment of customer demand is the most important, the second solution can be selected. The last solution will be the best choice if the decision-makers are concerned most about the overall cost.

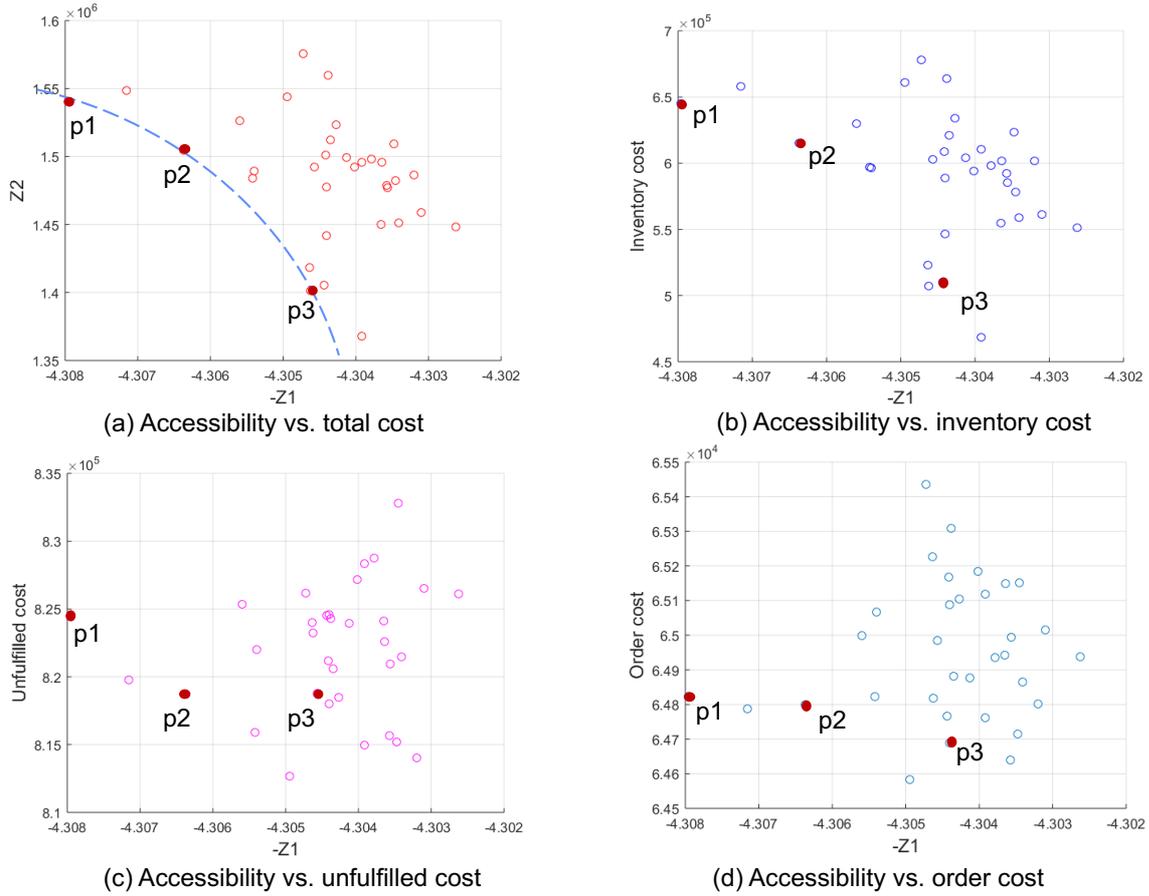

**Fig. 5.** Solutions on the Pareto front with large $Z_1$ and small $Z_2$ without considering safety stock: (a) accessibility vs. total cost; (b) accessibility vs. inventory cost; (c) accessibility vs. unfulfilled demand cost; (d) accessibility vs. order cost.

In the second experiment, we take the safety stock into consideration. We assume each DC should cover at least 40% of the local demand, and thus $v = 0.4$. Solutions are shown in Fig. 6. We can find there are three Pareto solutions. The first solution (p1) leads to the highest food accessibility but causes the highest overall cost. It would be the best choice if the decision-makers are concerned the most about food accessibility. The second solution (p2) has moderate food accessibility and cost but leads to the least inventory cost. The last solution (p3) has the least overall cost and the least unfulfilled demand cost. Therefore, if decision-makers are concerned most about the cost, the last solution can be the best choice.

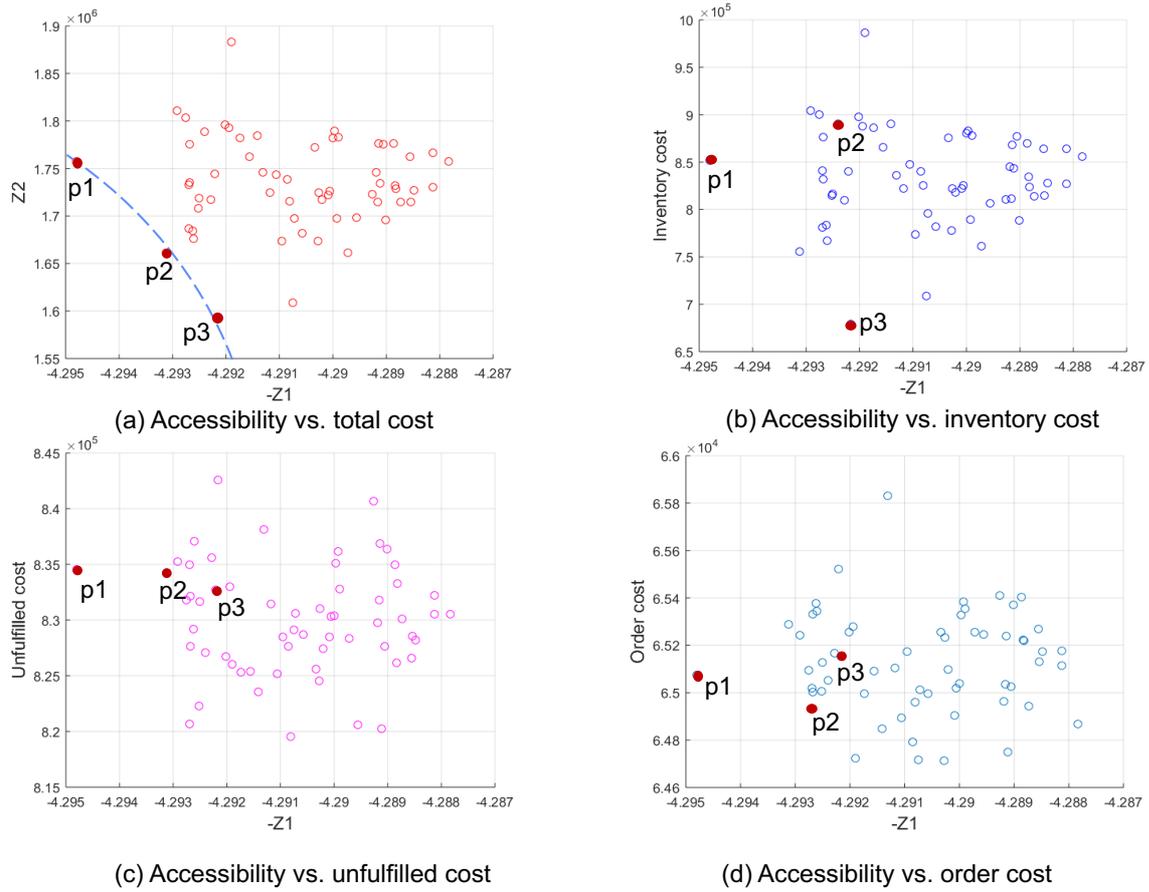

**Fig. 6.** Solutions on the Pareto front with large $Z_1$ and small $Z_2$ when $v = 0.4$. (a) accessibility vs. total cost; (b) accessibility vs. inventory cost; (c) accessibility vs. unfulfilled demand cost; (d) accessibility vs. order cost.

We compare the results of two experiments in Table 7. The first three rows display the results of two objective functions and three types of costs for solutions without safety stock. The last three rows show those for solutions considering safety stock. From the table, we can observe that when considering safety stock, food accessibility will have a small decrease. Because of the limited capacity $Cap_h^{DC}$, part of the space is occupied by safety stock. The overall cost with safety stock is higher because of the increased inventory cost. But we cannot ignore that safety stock help to cope with emergencies which we will consider in the future. Therefore, when making a choice, we should consider the trade-off between the two approaches.

**Table 7**

Comparison of Pareto solutions with safety stock and without safety stock (cost in 1000 QAR).

| Solutions | | $Z_1$ | $Z_2$ | Inventory cost | Unfulfilled demand cost | Order cost |
|---|---|---|---|---|---|---|
| **Without safety stock** | 1 | 4.3080 | 1540.47 | 645.17 | 824.44 | 64.82 |
| | 2 | 4.3046 | 1401.18 | 507.01 | 823.26 | 64.82 |
|                   | 3 | 4.3064 | 1504.47 | 614.94 | 818.71 | 64.80 |
|                   | 1 | 4.2931 | 1661.07 | 755.33 | 834.23 | 65.29 |
| **With safety stock** | 2 | 4.2922 | 1592.70 | 678.76 | 842.57 | 65.15 |
|                   | 3 | 4.2948 | 1757.31 | 851.47 | 834.57 | 65.08 |

*5.3. Solution evaluation in simulation*

When choosing from the three selected Pareto solutions, decision-makers should consider their pros and cons. In this section, we provide one possible way to help validate and evaluate the selected solutions: discrete event simulation. Simulation is able to implement order fulfillment strategies or demand patterns other than those considered in the mathematical formulation, thus providing validation and evaluation of the selected solutions under various more realistic scenarios.

We assume that in our proposed framework, at the beginning of each time period, we decide the distribution amount to each customer at the same time, so that every order can be processed (there is no dropped order) but may not be fulfilled. More specifically, each customer places $\widetilde{D}_{lt}$ order in period $t$, we assign $c_{hlt}$ product to each customer. We do not drop any orders but $c_{hlt}$ can be less than $\widetilde{D}_{lt}$ due to limited capacity. While this is reasonable, in another realistic scenario, orders may arrive at a random time. The simulation provides another realistic order fulfillment strategy. Once an order comes, the DC checks if the inventory can cover this order. If so, the order will be fully satisfied. If the current inventory cannot meet the order demand, this order will be dropped or wait until replenishment. Therefore, from the simulation, we can evaluate the three solutions from another realistic scenario.

For the three Pareto solutions on the POF, we put them in simulation and record the inventory cost, the unfulfilled demand cost, and the order cost respectively. In simulation setup, the customer demand is set as normal distribution $N(\mu, \sigma^2)$ with $\mu = 560$ and $\sigma^2 = 50$ as well. The total time period is set as $T = 5$. The policy is set as (S, s) policy and the safety stock s is set as $v \cdot S_h^{DC}$ for each DC. The initial stock is set to be $Inv_{h,0}^{DC}$ given by the Pareto solution.

The inventory cost, unfulfilled demand cost, order cost, and total cost are used to validate the results in our proposed framework. Since the simulation model mimics reality, if there is no significant difference between costs from those two approaches, our proposed framework is approved to be effective. On the other hand, due to different fulfillment strategies, we expect to see higher unfulfillment demand cost in simulation. Besides cost, we want to directly see what proportion of demands are met, because the satisfaction of customers is also important.

Service level is a performance metric scaled within [0,1] so that it can compare demand satisfaction more straightforwardly. For example, if total customer demands are met, the service level will be 1. Service level shows the percentage that customer demand is satisfied, which is calculated by Equation (28). Unsuccessful orders are defined by the placed orders requiring the number of products that is not available at the facility at the time when this order is placed. For example, if an order of 700 units of product comes to a DC, but the inventory is less than 700 units, this order will become an unsuccessful order.

$$\text{Service Level (by Products)} = \frac{\text{product in the successful orders}}{\text{sum of products in all orders placed for this facility}} \quad (28)$$

Table 8 shows the comparison of performance metrics given by the three Pareto solutions between our proposed framework and the simulation model. It can be seen that the performances given by the two approaches are quite consistent. The inventory cost and order cost are quite similar, except for some differences caused by the randomness in the distance. The unfulfilled demand cost from the simulation is higher than that of the proposed framework. This is likely to be caused by the simulation logic that the DC does not fulfill a partial order. If there is not enough inventory when an order comes in, the order has to wait until DC's replenishment, resulting in more unfulfillment. This agrees with our conjecture. To take a closer look at the service level in each region separately, we summarize the service level in each of the 4 regions. We find that Region 1 has the lowest service level because it has a large number of customers, making fulfilling all orders more difficult. Service levels in other regions are higher because it is relatively easier to fulfill the demand for a smaller number of customers.

Although this discrete event simulation uses a different ordering rule from the optimization model, it provides a different perspective and helps to evaluate our solutions in a more realistic environment. Results on the service level show how the candidate solutions can perform in another realistic scenario and give a reference for solution selection, e.g., we may recommend solution p3 if we are more concerned with the service level.

**Table 8**

Comparison of performance metrics between the proposed framework and simulation model (cost in 1000 QAR).

| Method | Cost | Pareto efficient solutions | | |
| --- | --- | --- | --- | --- |
| | | p1 | p2 | p3 |
| **Proposed framework** | Inventory cost | 755.33 | 678.76 | 851.47 |
| | Unfulfilled demand cost | 834.23 | 842.57 | 834.57 |
| | Order cost | 65.29 | 65.15 | 65.08 |
| | Total cost | 1661.07 | 1592.70 | 1757.31 |
| **Simulation** | Inventory cost | 751.84 | 667.65 | 969.39 |
|  | Unfulfilled demand cost | 953.38 | 995.01 | 948.32 |
|  | Order cost | 72.24 | 66.96 | 62.95 |
|  | Total cost | 1777.46 | 1729.61 | 1980.67 |
|  | Region 1 | 0.428 | 0.424 | 0.443 |
|  | Region 2 | 0.942 | 0.924 | 0.940 |
| **Service level** | Region 3 | 0.848 | 0.851 | 0.901 |
|  | Region 4 | 1 | 1 | 1 |
|  | Total | 0.590 | 0.585 | 0.603 |

## 6. Conclusion and future work

In this paper, we develop an optimization model to determine the optimal design for food supply chains with two objectives, food accessibility and cost. A food accessibility evaluation index system is proposed to measure local food accessibility. With the presence of random supply and demand, our problem is formulated as a stochastic multi-objective mixed-integer programming problem with nonlinear constraints. To solve this problem, we propose a two-phase framework in which long-term, tactical decisions and short-term, and operational decisions are determined systematically. In Phase I, we determine the supply chain configuration, so that integer variables in the optimization model are solved. In Phase II, continuous variables representing supply chain operations are solved using a Monte Carlo simulation approach. A set of feasible solutions are obtained by iteratively performing a Monte Carlo simulation. Pareto efficient solutions are selected from the Pareto front and validated separately in discrete event simulation to evaluate the performance of the designed supply chain in a more realistic scenario and provide recommendations for different decision-makers.

Future work can be expanded in three directions. First, the results can be improved with more accurate data. In our experiment, parameters in optimization and simulation models are decided by historical data from relevant papers and open resources. Future work will look into approaches to improving the availability of real-time data. For example, we may consider developing an IoT database using real-time data to determine parameters in our model, so that the data can be updated, and optimal solutions can be determined accordingly. Second, considering that the income, the population, and the price are fluctuating, it could be more beneficial to integrate our proposed framework with forecasting models to predict parameters (e.g., future income, future food price, and future population), which helps to make a response in advance. Third, we are interested in using our model to give decision support for supply chain network design under disruption caused by emergency scenarios. We believe in its potential in improving supply chain resilience and facilitating accurate decision-making and in-time adjustment for the supply chain network design.

## Funding

This study is supported by the Qatar National Research Fund (a member of Qatar Foundation) under Grant Award Number MME02-1004-200041.